\documentclass[twocolumn, trackchanges, floatfix]{aastex631}

\usepackage{amsmath}
\usepackage{amssymb}
\usepackage{bm}
\usepackage{booktabs}
\usepackage{multirow}

\begin{document}

\title{The effect of intrinsic alignments on weak lensing statistics in hydrodynamical simulations}

\correspondingauthor{Max E. Lee}
\email{max.e.lee@columbia.edu}

\author[0000-0002-0786-7307]{Max E. Lee}
\affiliation{Department of Astronomy, Columbia University, MC 5246, 538 West 120th Street, New York, NY 10027, USA}
%\affiliation{Columbia University Department of Astronomy, 538 West 120th Street, New York, NY 10027, USA}

\author[0000-0003-3633-5403]{Zolt\'an Haiman}
\affiliation{Department of Astronomy, Columbia University, MC 5246, 538 West 120th Street, New York, NY 10027, USA}
\affiliation{Department of Physics, Columbia University, MC 5255, 538 West 120th Street, New York, NY 10027, USA}
\affiliation{Institute of Science and Technology Austria, Am Campus 1, Klosterneuburg 3400 Austria}
% \affiliation{American Astronomical Society \\
% 1667 K Street NW, Suite 800 \\
% Washington, DC 2,0006, USA}

% \collaboration{20}{(AAS Journals Data Editors)}

\author{Shivam Pandey}
\affiliation{William H. Miller III Department of Physics \& Astronomy,
Johns Hopkins University, Baltimore, MD 21218, USA.}
\affiliation{Columbia Astrophysics Laboratory, Columbia University, 550 West 120th Street, New York, NY 10027, USA}

\author[0000-0002-3185-1540]{Shy Genel}
\affiliation{Center for Computational Astrophysics, Flatiron Institute, 162 Fifth Ave, New York, NY, 10010, USA}
\affiliation{Columbia Astrophysics Laboratory, Columbia University, 550 West 120th Street, New York, NY 10027, USA}
% \altaffiliation{AASTeX v6+ programmer}
% \affiliation{TeXnology Inc.}



\begin{abstract}
The next generation of weak gravitational lensing surveys has the potential to place stringent constraints on cosmological parameters. However, their analysis is limited by systematics such as the intrinsic alignments of galaxies, which alter weak lensing convergence and can lead to biases in cosmological parameter estimations. In this work, we investigate the impact of intrinsic alignments on non-Gaussian statistics of the weak lensing field using galaxy shapes derived from the IllustrisTNG hydrodynamical simulation. We create two catalogs of ray-traced convergence maps: one that includes the measured intrinsic shape of each galaxy and another where all galaxies are randomly rotated to eliminate intrinsic alignments. We compare an exhaustive list of weak lensing statistics between the two catalogs, including the shear-shear correlation function, the map-level angular power spectrum, one-point, peak count, minimum distribution functions, and Minkowski functionals. For each statistic, we assess the level of statistical distinguishability between catalogs for a set of future survey angular areas. Our results reveal strong small-scale correlation in the alignment of galaxies and statistically significant boosts in weak lensing convergence in both positive and negative directions for high-significance peaks and minimums, respectively. Weak lensing analyses utilizing non-Gaussian statistics must account for intrinsic alignments to avoid significantly compromised cosmological inferences.
\end{abstract}


% \keywords{}


\section{Introduction} \label{sec:intro}
The next generation of cosmological surveys (Stage IV), such as the Vera Rubin Legacy of Space and Time telescope (LSST; \citealt{LSST}), the Nancy Grace Roman space observatory (Roman; \citealt{Roman15}), and the Euclid Mission (Euclid; \citealt{Euclid11}) can provide robust tests of the standard cosmological model, $\Lambda$CDM, and offer insights into the current tensions among its parameters (See \citealt{Perovolaropoulos-2022} for a comprehensive guide to current $\Lambda$CDM tensions). 

Weak gravitational lensing (WL) is a particularly promising probe of the cosmological model because of its ability to probe parameters at the forefront of these tensions: the matter density of the Universe, the amplitude of matter fluctuations, and the dark energy equation of state ($\Omega_M$, $\sigma_8$, $w$) \citep{Hikage-2019, Hamana-2020, DES21}. With areas up to $18,000\,{\rm deg}^2$ covered, lens redshifts of up to $z\sim 2$, a galaxy population of $\sim 50\, {\rm arcmin}^{-2}$, and the ability to observe galaxies with magnitudes $\leq 26.5$ \citep{LSST, Mahony-2022}, the analysis of the measurements will be limited by uncertainties on the smallest scales arising from systematic effects. This includes complex baryonic physics \citep{Jing06, Chisari19}, photometric redshift errors \citep{Bernstein-2010}, and correlations between the intrinsic shapes of galaxies, frequently referred to as intrinsic alignments (IA) \citep{Schndeider-2010, Sifon-2015}. Our understanding and the level of control over these modeling systematics will ultimately dictate the precision of the cosmological information such surveys can extract from WL.

There has been a significant push in WL research to better understand the effect of systematics on statistics extracted from the WL field and on the resulting cosmological parameter constraints \citep{Blazek-2015, Schneider-2019, Arico-2020, Vlah-2020, Arico-2021, Lee-2023}. Baryonic effects are expected to suppress power at intermediate scales while enhancing power at the smallest scales for two-point statistics like the power spectrum and correlation function \citep{Schneider-2019, Arico-2021, Osato-2021}. Similarly, studies consistently find correlations between galaxy shapes at small scales, causing an enhancement in the power spectrum \citep{Chisari-2015, delgadoMillenniumTNGProjectIntrinsic2023}. 

Statistics beyond traditional two-point functions such as the lensing three-point function and bispectrum \citep{Takada-2003}, weak lensing peak and minimum counts \citep{Davies-2022}, and Minkowski functionals \citep{Marques-2019} have been shown to contain a wealth of information and are typically called non-Gaussian statistics due to their ability to probe the non-Gaussian nature of the weak lensing field. However, modeling the influence of systematic effects in higher-order and non-Gaussian statistics is complex and challenging to untangle. 

Previous work has considered IA for non-Gaussian statistics, such as \cite{zhangEffectsGalaxyIntrinsic2022}, where the authors investigate peak counts using a semi-analytical galaxy formation model. They find that when satellite galaxies are radially aligned with their host galaxy, peaks can be suppressed by up to $75\%$ compared to peaks generated from purely random galaxy alignments. Prescribing IAs to galaxies as done in \cite{zhangEffectsGalaxyIntrinsic2022} is an efficient approach to exploring scenarios where IA must be considered. However, this prescriptive approach neglects the complexities of alignments that arise naturally due to galaxy formation and evolution. Probing an IA signal that includes galaxy formation and evolution requires using galaxies extracted directly from hydrodynamical simulations, where the shapes, luminosities, and orientations of galaxies are expected to resemble reality more closely than semi-analytical prescriptions. To date, no study has investigated the impact of IAs on non-Gaussian statistics following this approach. 

In this work, we perform the first analysis of non-Gaussian statistics of the weak lensing field that includes IAs computed directly from galaxies in the IllustrisTNG magneto-hydrodynamical simulations. We explore the influence of IAs on various two-point and non-Gaussian statistics, including the shear-shear correlation function, angular power spectrum, one-point probability, peak count, minimum distribution functions, and three Minkowski functionals. We compare the effects of IA on these statistics and identify the survey area at which the effect of IA becomes distinguishable from the typically assumed random orientation. 

The structure of this paper is as follows. In \S~\ref{sec:background}, we provide a brief background on weak gravitational lensing, IAs, the IllustrisTNG simulation suite, and ray-tracing. We then detail our analysis pipeline in \S~\ref{sec:pipeline}. We present and explore our results on the various statistics in \S~\ref{sec:stats}, and conclude with some statements on current and planned surveys in \S~\ref{sec:conclusion}.

\section{Background}\label{sec:background}
In this section, we first briefly review the theoretical framework for weak gravitational lensing and the intrinsic alignment of galaxies (see \citealt{Bartelmann-2001, Hoekstra-2008, Kilbinger-2015} for reviews on weak gravitational lensing, and \citealt{Lamman-2023} and references within for reviews on the IA of galaxies). We then describe the hydrodynamical simulations and the ray-tracing procedure used in this work.

\subsection{Weak Lensing}
A light ray's path from a source galaxy at its true sky location to an observer at the origin is deflected by intervening matter. For small distortions to the light ray's path, the Jacobian describing the mapping from the angular position of the true source to the observed source is given by,
\begin{equation}\label{eq:A}
    \begin{split}
        \bm{A} = \begin{pmatrix}
            1-\bm{\kappa}- \bm{\gamma}_1\ \ &-\bm{\gamma}_2\\
            -\bm{\gamma}_2\ \ &1 - \bm{\kappa}+ \bm{\gamma}_1
        \end{pmatrix}
    \end{split}
\end{equation}
where the \textit{convergence}, $\bm{\kappa}$ and complex \textit{gravitational shear} $\bm{\gamma}=\bm{\gamma}_1+ i\bm{\gamma}_2$  are related to the two-dimensional lensing potential, $\psi$ as,
\begin{equation}\label{eq:kappa_gamma}
    \begin{split}
        \bm{\kappa}&= \dfrac{1}{2} \nabla^2\bm{\psi}\\
        \bm{\gamma}_1&= \dfrac{1}{2}\dfrac{\partial^2 \bm{\psi}}{\partial \theta_1^2} -\dfrac{1}{2}\dfrac{\partial^2 \bm{\psi}}{\partial \theta_2^2}\\
        \bm{\gamma}_2&= \dfrac{\partial^2\bm{\psi}}{\partial\theta_1\partial\theta_2}.
    \end{split}
\end{equation}
Here, $\theta_1$ and $\theta_2$ are the angular sky coordinates, making up the vector $\bm{\theta}$. The two-dimensional lensing potential is proportional to the line of sight integral of the gravitational potential, $\Phi$, between the observer and the source redshift,
\begin{equation}
    \bm{\psi}(\bm{\theta}) = \dfrac{D_{ls}}{D_lD_s}\dfrac{2}{c^2}\int \bm{\Phi}(D_l\bm{\theta}, z)dz.
\end{equation}
The prefactor is the ratio of the distances between the source galaxy and lens, $D_{ls}$, to the distances to the lens, $D_l$, and the source, $D_s$. The matter density field can be related directly to the potential through the Poisson equation, 
\begin{equation}\label{eq:poisson}
    \nabla^2\bm{\Phi} = 4\pi G\bm{\rho}.
\end{equation}
\indent This highlights that the convergence field, $\bm{\kappa}$, in conjunction with Eq.~\ref{eq:poisson}, represents a projected matter distribution along the line of sight and is, therefore, a sensitive probe of cosmological parameters that influence this distribution. 

The convergence can be extracted from simulations and observations. In simulations, $\bm{\kappa}$ can be computed either through approximations (such as the Born approximation) or more directly through ray-tracing of light paths between objects and the origin (see \S~\ref{sec:ray-tracing}, and \citealt{Petri-2017} or \citealt{Ferlito-2024} for a comparison between methods).

The WL field is found for galaxy surveys from observations of projected galaxy ellipticities \citep{Abbot-2022, Miyatake-2023}. The relationship between observed ellipticities and $\bm{\kappa}$ or $\bm{\gamma}$ is through the definition of observed ellipticity as follows: (e.g. \citealt{Seitz-1997}):
\begin{equation}\label{eq:ellipticity}
    \begin{split}
        \bm{\epsilon}  &= \begin{cases}
            \dfrac{\bm{\epsilon_s} + \bm{g}}{1 + \bm{g}^\ast\bm{\epsilon_s}} \ \ \ &|g|\leq 1\\
            \dfrac{1 + \bm{g}\bm{\epsilon_s}}{\bm{\epsilon_s}^\ast + \bm{g}^\ast} \ \ \ &|g| > 1\\
        \end{cases}\\
    \bm{g} &= \dfrac{\bm{\gamma}}{1-\bm{\kappa}},
    \end{split}
\end{equation}
where $\bm{g}$ is the \textit{reduced shear}, and $\bm{\epsilon}_s$ is the intrinsic projected ellipticity of the observed galaxy. 

Observations can only measure the quantity $\bm{\epsilon}$ in Eq.~\ref{eq:ellipticity}. However, if the intrinsic orientations of the galaxies are all random, then the expectation value of the observed ellipticity over an infinite number of galaxies is to first order the reduced shear, 
\begin{equation}\label{eq:ellipt_to_gamma}
    \langle\bm{\epsilon}\rangle \sim \langle \bm{g}\rangle.
\end{equation}
\indent For small $\bm{\kappa}$, the quantity $\bm{g}$ can be approximated as $\bm{g}\simeq\bm{\gamma}$, such that the ellipticity becomes an estimator for the shear. 
\begin{figure*}
    \centering
    \includegraphics[width=\linewidth]{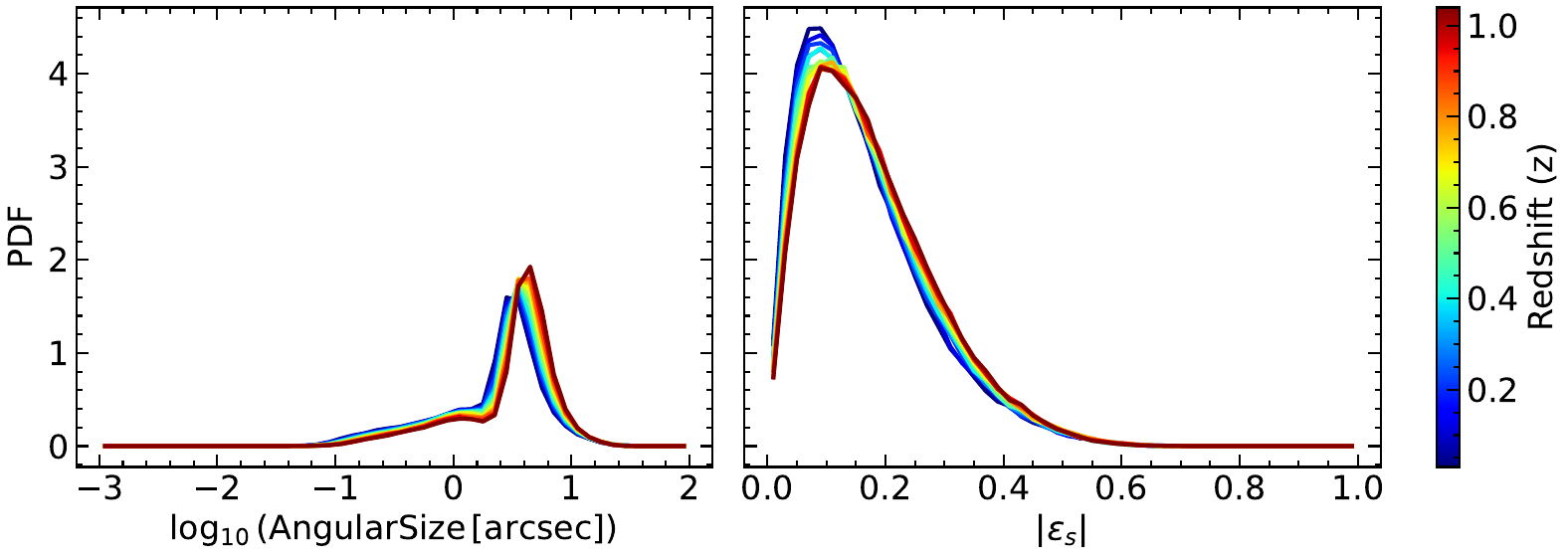}
    \caption{We show the subhalo angular size and ellipticity distributions for our selected galaxies at individual redshift slices up to redshift $z=1.04$. Most galaxies have sub-arcsec angular sizes and low ellipticity, and we find a moderate evolution of both ellipticity and angular size with respect to redshift.}
    \label{fig:galaxy_char}
\end{figure*}
The relationship between $\bm{\kappa}$ and $\bm{\gamma}$ in Eq.~\ref{eq:kappa_gamma} can be solved in Fourier space, leading to the commonly used Kaiser-Squires inversion \citep{kaiserMappingDarkMatter1993, Seitz-1997}
\begin{equation}\label{eq:KS}
    \tilde{\bm{\gamma}}(\bm{k}) = \dfrac{1}{\pi} \tilde{\bm{D}}(\bm{k})\tilde{\bm{\kappa}}(\bm{k})
\end{equation}
where
\begin{equation}
    \Tilde{\bm{D}}(\bm{k}) = \pi\dfrac{k_1^2 - k_2^2 + 2ik_1k_2}{k_1^2 + k_2^2},
\end{equation}
the $\tilde{.}$ denotes a Fourier transformation and $\bm{k}$ is the two-component scale or Fourier dual to $\bm{\theta}$.

In this work, we use both the simulation and observational approaches. We treat each galaxy in a simulation box as one in which we can measure its intrinsic ellipticity, $\bm{\epsilon}_s$, and the value of convergence and shear through ray-tracing. We can then find the impact of IAs on WL observables by computing the observed ellipticity of Eq.~\ref{eq:ellipticity}, converting to shear following Eq.~\ref{eq:ellipt_to_gamma}, and following the observational procedure of using Eq.~\ref{eq:KS} to derive mock convergence maps.

\subsection{Simulation suite}
We use the IllustrisTNG300-1 (hereafter TNG300) simulation \citep{Pillepich-2018, springel-2018, Nelson-2018, naiman-2018, Marinacci-2018}, a cosmological magneto-hydrodynamical simulation employing a wide range of physical processes that drive galaxy formation and evolution. The TNG300 simulation box covers a volume of $(205\,h^{-1}\,{\rm Mpc})^3$ with dark matter and gas particles with a mass resolution of $3.98\times10^7\,h^{-1}\,M_\odot$ and $7.44\times10^6\,h^{-1}\,M_\odot$, respectively. The cosmological model uses the results of Planck 2015 \citep{Planck-2016}: $\Omega_m=0.3089$, $\sigma_8=0.8158$, $\Omega_b=0.0486$, $H_0=67.74\,{\rm km}\,{\rm s}^{-1}\,{\rm Mpc}^{-1}$. Halo and subhalo catalogs are computed on the fly and saved for each redshift snapshot via the FoF/Subfind algorithms \citep{springel-2001, Nelson-2018, Pillepich-2018}. TNG300 contains 100 simulation snapshots and FoF/Subfind catalogs output between redshifts $z=0-20$, but for a WL study, we are primarily interested in redshifts between $0\leq z\leq1$, as these will be the redshift ranges probed by upcoming WL surveys.

We use the dark matter, gas, and stellar distributions from $12$ snapshots in TNG300 between $0.03\leq z\leq 1.04$ and associated FoF/Subfind catalogs of halos and subhalos. We limit our galaxy sample to only include resolved galaxies, which are those with more than 1,000 stellar particles \citep{Donnari-2021} and which have cosmological origins, meaning they formed due to the process of structure formation and collapse. Following these limits, our galaxy sample contains 1,188,366 galaxies in the redshift range. We show a brief characterization of this sample in Fig.~\ref{fig:galaxy_char}, highlighting the angular size and ellipticity distributions. 

The angular sizes of galaxies are typically small at subarcsec scales, which we find to be smaller than the cumulative deflections from cosmic shearing in \S~\ref{sec:ray-tracing}. We discuss the computation of ellipticities in the following section.

\subsection{Galaxy elipticities}
We assume each galaxy to be a triaxial ellipsoid with major, median, and minor axes $(\bm{a}, \bm{b}, \bm{c})$ corresponding to the eigenvectors from of the luminosity analog of the inertia tensor,
\begin{equation}\label{eq:SIT}
    \begin{split}
        \bm{I}_{ij} = \dfrac{1}{L}\sum_\alpha L_{\alpha}x_{i, \alpha}x_{j, \alpha}
    \end{split}
\end{equation}
where $L$ is the luminosity of each stellar particle, the galaxy's total luminosity is given by $L=\sum_\alpha L_\alpha$, $x_i, x_j$ are the positions of the stellar particles with respect to the center of the galaxy light.

We compute luminosities by extracting all particle positions, ages, metallicities, and masses from resolved subhalos. Using the stellar population synthesis code, \textit{FSPS}\footnote{\href{https://dfm.io/python-fsps/current/}{dfm.io/python-fsps}} (Flexible Stellar Population Synthesis; \citealt{Conroy-2009, Conroy-2010}), we compute the luminosity of each stellar particle in a given galaxy, emulating the observation of the galaxy. 
\begin{figure}
    \centering
    \includegraphics[width=0.98\linewidth]{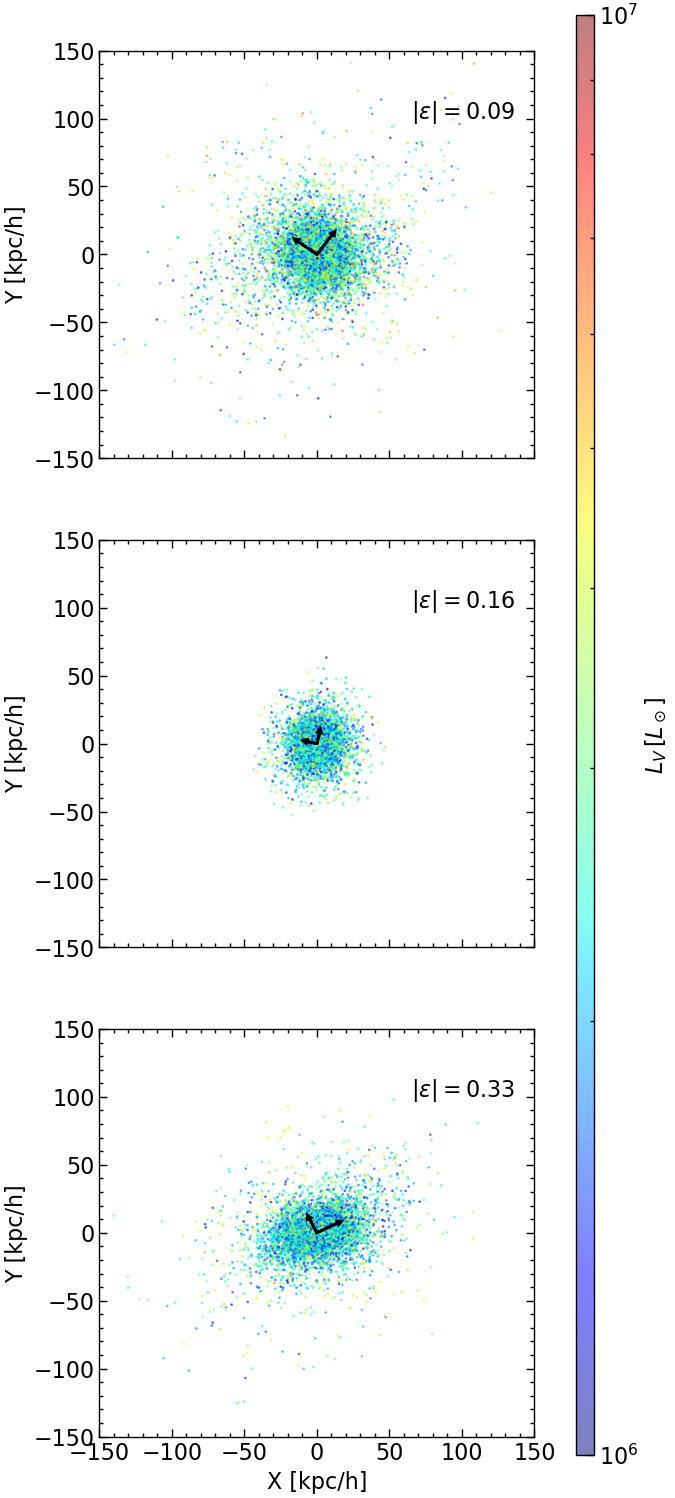}
\caption{We present the stellar component of three galaxies in our sample. For each galaxy, we plot the individual star particles associated with the galaxy, and we color each point according to the calculated V-band Luminosity. The luminosity and positions of the star particles are then used to find the galaxy's center and the brightness tensor's second moment (Eq.~\ref{eq:SIT}). The arrows represent the projected inertia tensor's scaled eigenvectors and the galaxy's principal axes. Finally, we show the magnitude of the galaxy's ellipticity for each galaxy.}
    \label{fig:galaxy_sample}
\end{figure}

The projected ellipticity when viewed along the $z$-axis (axis-3) is then given by the complex number, $\epsilon = \epsilon_1 + i\epsilon_2$ where
\begin{equation}\label{eq:ellipt}
    \begin{split}
        \epsilon_1 = \dfrac{I_{11}-I_{22}}{I_{11} + I_{22}}\\
        \epsilon_2 = \dfrac{2I_{12}}{I_{11} + I_{22}},
    \end{split}
\end{equation}
though the projection can occur in any direction by simply changing the appropriate indices above.

As discussed in \S~\ref{sec:ray-tracing}, we choose random axes to project along during ray-tracing to generate a range of pseudo-independent light-cones. However, we can always use equivalent equations to the above to find the correct projected ellipticity. 

As an example of this calculation, in Fig.~\ref{fig:galaxy_sample}, we show the stellar components of three galaxies in our sample at redshift $z=0.03$, where each point represents the absolute magnitude of a star particle in that galaxy. To find the major and minor axes in the projected ellipse, we compute the eigenvalues and eigenvectors of the 2D luminosity inertia tensor given by,
\begin{equation}
    I_{xy} = \begin{pmatrix}
        &I_{11} \ \ \ &I_{12}\\
        &I_{21} \ \ \ &I_{22}
    \end{pmatrix}.
\end{equation}
We show the axes of each galaxy following this calculation as arrows by computing the eigenvectors, $\bm{e}=(\sqrt{\lambda_1}\bm{v}_1, \sqrt{\lambda_2}\bm{v}_2)$. We also show the magnitude of the ellipticity vector following Eq.~\ref{eq:ellipt}.

In Fig.~\ref{fig:galaxy_sample}, we see the orthogonal axes describing the two components of the ellipticity. The lower panel, which has a clear elongation, has the highest magnitude of ellipticity of this subset, while the top panel appears the most circular. Referring back to Fig.~\ref{fig:galaxy_char}, few galaxies in our sample are purely circular (with $|\epsilon|=0)$, and, as expected, none have ellipticities of $|\epsilon|=1$. The PDF of ellipticity changes with galaxy redshift on an individual snapshot level. While there is not a massive change with redshift, galaxies tend to become more circular at lower redshifts. This result and our ellipticity distributions are concordant with \cite{samuroff_advances_2021}, who performed a similar ellipticity calculation with TNG100. 

\section{Pipeline}\label{sec:pipeline}
\begin{figure*}
    \centering
    \includegraphics[width=\linewidth]{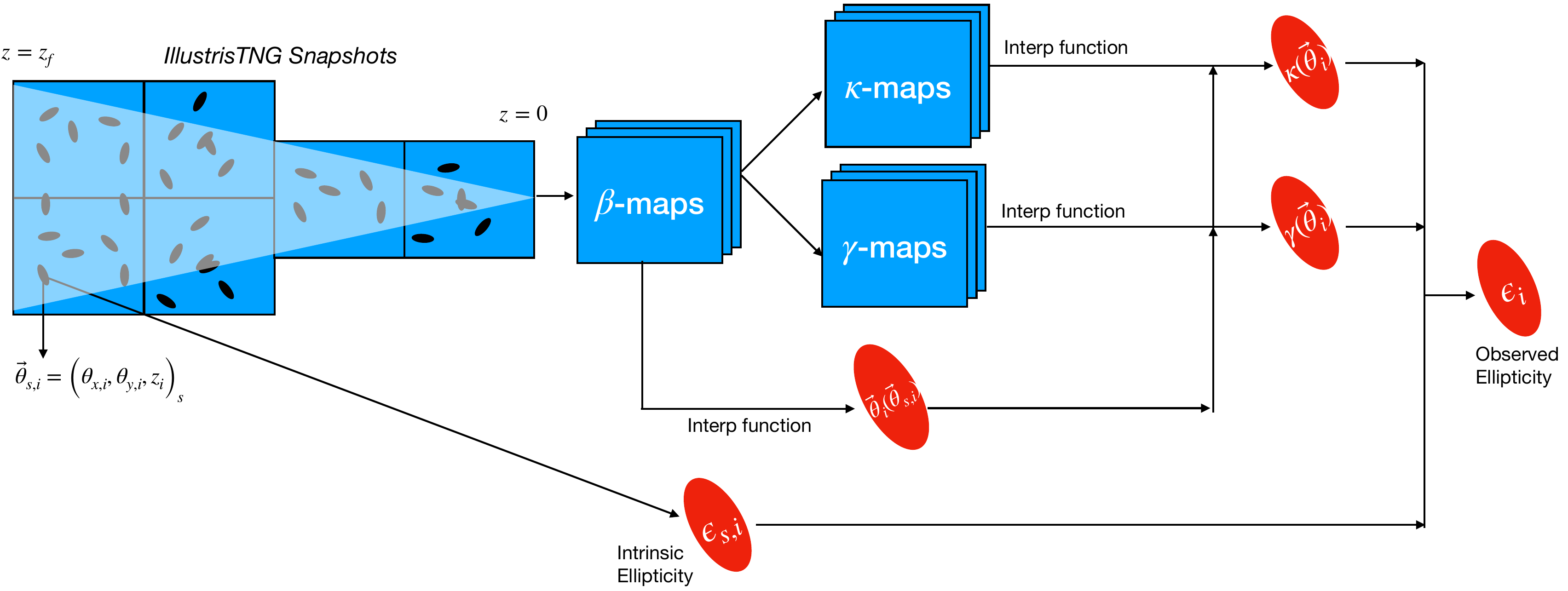}
    \caption{Illustration of deriving observed ellipticities from TNG300 that include IA measurements from each galaxy. The first step is to extract the observed ellipticity of each galaxy, $\epsilon_i$, by finding the convergence $\kappa_i$, shear $\gamma_i$, and intrinsic ellipticity $\epsilon_{s,i}$. We illustrate the broad process to perform this. First, we generate a light-cone from redshift $z=0$ to some final redshift by tiling together TNG300 snapshots. We perform ray-tracing to which following Eq.~\ref{eq:beta} finds the angular derivatives of the light-ray deflections, $\bm{\beta}$, to generate $\bm{\kappa}$ (convergence), $\bm{\gamma}$ (shear) maps at several redshifts up to $z_f$. For each source galaxy, $i$, with intrinsic angular position $\vec{\theta}_{s,i}$, we find the observed position, $\vec{\theta}_i$ by interpolating the $\bm{\beta}$ maps at the true location. We then interpolate the $\bm{\kappa}$ and $\bm{\gamma}$ maps to find the $\bm{\kappa}$ and $\bm{\gamma}$ values for the galaxy at the observed position. Finally, we measure the galaxy's intrinsic ellipticity, $\epsilon_{s,i}$, and combine this with the $\bm{\kappa}$ and $\bm{\gamma}$ values to find the galaxy's observed ellipticity.}
    \label{fig:flowchart}
\end{figure*}
\indent Generating TNG300 convergence maps with the influence of IA requires that we 1. find values for $\bm{\kappa}$ and $\bm{\gamma}$ at each galaxy's observed position, 2. find each galaxy's projected intrinsic ellipticity, 3. obtain the simulated observed ellipticity of Eq.~\ref{eq:ellipticity} and observed shear of Eq.~\ref{eq:ellipt_to_gamma}, and 4. convert the shear to convergence via the Kaiser-Squires inversion of Eq.~\ref{eq:KS}. We schematically illustrate this procedure in Fig.~\ref{fig:flowchart} and describe each step in detail in the following section. 

\subsection{Ray-tracing} \label{sec:ray-tracing}
We follow the procedure of \cite{Osato-2021} (henceforth $\bm{\kappa}$TNG) and, more generally, the multiple lens-plane algorithm (ray-tracing) \citep{Jain-2000, Hilbert-2009, Petri-2016a} to generate convergence and shear maps from the TNG300 simulation. $\bm{\kappa}$TNG stacks twelve TNG300 snapshots between $0\lesssim z\lesssim1$ to generate light-cones of $25\,{\rm deg}^2$. Each box is randomly rotated and translated to remove the overlap of structures in a given light-cone and to allow for the generation of numerous pseudo-independent light-cones (see \citealt{Petri-2016} for more details).

We ray-trace each light-cone, approximating the matter distribution as density planes with thickness $\delta\chi$ and surface density $\sigma$. Simulation boxes are broken into two lens-planes with $\delta\chi=102.5\,h^{-1}\,{\rm Mpc}$, and the lensing potential at each plane, $k$, is found with
\begin{equation}
    \nabla^2_{\bm{\beta}^k} \psi^k(\bm{\beta}^k) = 2\sigma^k(\bm{\beta}^k),
\end{equation}
where $\bm{\beta}^k$ is the angular position on the $k^{\rm th}$ lens-plane. For a given light ray, its path travels in straight lines between lens-planes, and is deflected by angle $\bm{\alpha}$, which is related to the lensing potential as 
\begin{equation}
    \nabla_{\bm{\beta}^k} \psi^k(\bm{\beta}^k) = \alpha^k(\bm{\beta}^k).
\end{equation}
When the light ray reaches its final (source) redshift, the final lensed position of the ray represents the sum of the previous deflection angles,
\begin{equation}\label{eq:beta}
    \bm{\beta}^k(\bm{\theta}) = \bm{\theta} - \sum_{i=1}^{k-1}\dfrac{\chi^k-\chi^i}{\chi^k}\bm{\alpha}^i(\beta^i), (k=2, 3,\dots)
\end{equation}
The lensed position is used to reconstruct the lensing Jacobian matrix of Eq.~\ref{eq:A}, 
\begin{equation}
    A_{ij}(\bm{\theta}, \chi) \equiv \dfrac{\partial\beta_i(\bm{\theta}, \chi)}{\partial\theta_j}.
\end{equation}
For each lens-plane, we follow the same procedure as $\bm{\kappa}$TNG, projecting the dark, stellar, and gaseous matter of TNG300 within slices of $\delta\chi=102.5\,h^{-1}\, {\rm Mpc}$ to a pixelized grid with $4096^2$ pixels using triangular-shaped cloud interpolation. Note, that this amounts to two lensplanes generated per TNG300 snapshot used, each of which contains the same matter distribution, though they are assigned different redshifts in the lightcone. At each lens-plane and for each pixel, we compute the deflection angle, which we use to find $1024^2$ pixel grids of $\bm{\beta}$, $\bm{\kappa}$, and $\bm{\gamma}$. 
\begin{figure*}
    \centering
    \includegraphics[width=\linewidth]{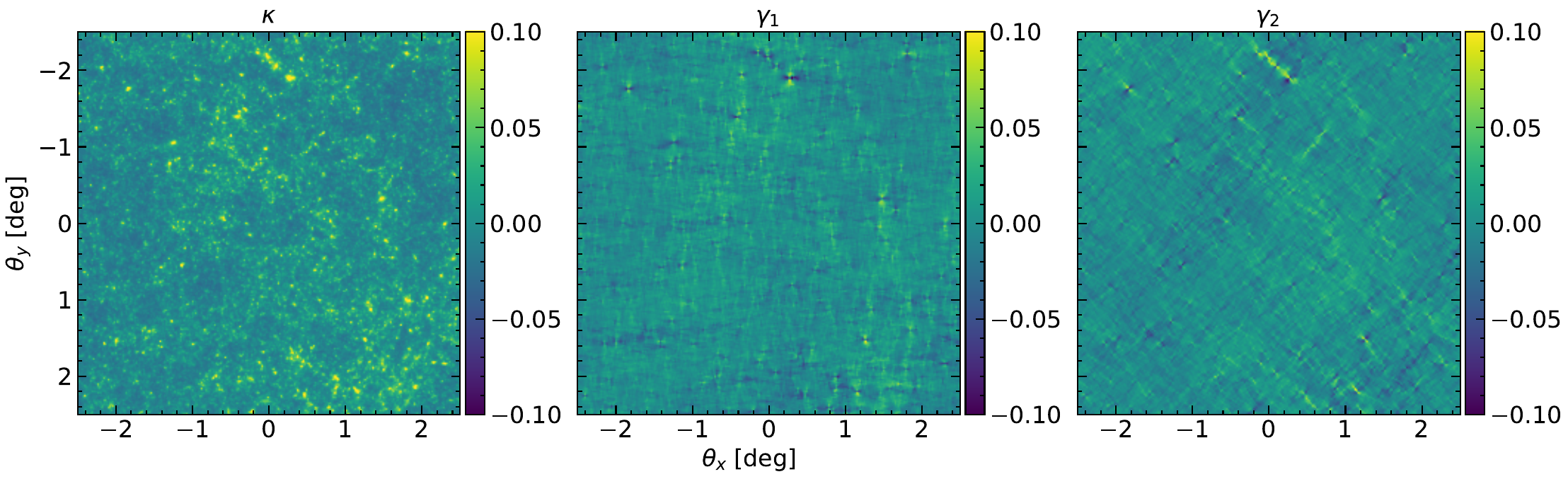}
    \caption{A typical ray-tracing output upto $z_{\rm max}=1$, using the ray-tracing methodology described in Sec.~\ref{sec:ray-tracing}. The left panel shows the convergence, or $\bm{\kappa}$ map, while the center and right panel show the two components of the shear or $\bm{\gamma}$.}
    \vspace*{5mm}
    \label{fig:kappa_gamma}
\end{figure*}

\indent In Fig.~\ref{fig:kappa_gamma}, we show the results of ray-tracing up to a source redshift of $z=1$. The left-most plot shows the convergence or $\bm{\kappa}$ map, while the right two plots show the two components of the shear or $\bm{\gamma}$ maps. 

The $\bm{\beta}$ maps play a fundamental role in generating the convergence and shear maps but also allow us to transform the intrinsic positions of galaxies in TNG300 to the observed positions. This is critical to appropriately match $\bm{\kappa}$ and $\bm{\gamma}$ values to each galaxy. 
\begin{figure}
    \centering
    \includegraphics[width=\linewidth]{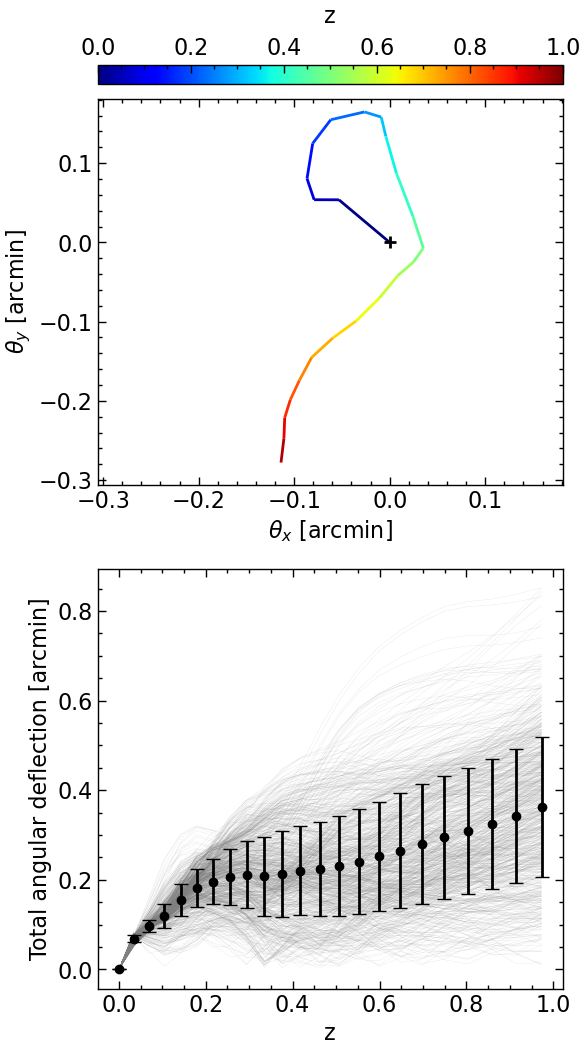}
    \caption{As light propagates through lens-planes, its angular position shifts around by arcsecond scales per mass plane, culminating in a sub-arcminute net displacement. In the top panel we show the path of a single lightray with initial position $\theta_x,\theta_y=(0,0)$ at $z=0$ and the path of deflections up to its final position at redshift $z=1$. We mark the origin with a $+$ to guide the eye. The bottom pannel shows the distance travelled between initial and final positions as a function of redshift, $d(z)=\sqrt{\Delta \theta_x(z)^2 + \Delta\theta_y(z)^2}$, for $40,000$ randomly chosen pixels along with the mean and standard deviation over this subset.}
    \label{fig:beta_map}
\end{figure}
In Fig.~\ref{fig:beta_map}, we show the path of a light ray from its initial position at $z=0$ to the deflected final position at $z=1$. Each lens-plane induces a sub-arcminute deflection to the light ray, culminating in a total displacement of $\sim 0.36\,\rm arcmin$ between initial and final positions. We show in the bottom panel a set of $40,000$ such paths chosen randomly from the set of $1024^2$ pixels. We present the distance of the paths as a function of redshift, and highlight that the mean displacement between initial and final positions is $\sim0.36\pm0.15\,\rm arcmin$. The typical angular size of a galaxy at a redshift of $z=1$ is less than an arcsecond (as shown in Fig.~\ref{fig:galaxy_char}), which means that the lensing effect can show significant differences between observed and intrinsic galaxy positions.

We generate $2,000$ unique light-cones by randomly rotating the simulation boxes either by $0\deg, 90\deg, 180\deg$ or $270\deg$ about each of the three axes, translating the particles in each of the directions, and choosing an axis for projection. The projected density field is then used to compute the first and second derivatives of the potential. We then perform an additional random rotation of $90\deg$ or $270\deg$ (See \href{sec:flip}{Appendix A} for details on this choice) and translation of each lens-plane while tracing lightpaths to find $\bm{\kappa}$ and $\bm{\gamma}$. This rotation and translation process allows for a large statistical sample of WL map realizations and has been shown to yield up to 10,000 statistically indistinguishable realizations in \cite{Petri-2016}.
\subsection{Aligning galaxies}
\begin{figure*}
    \centering
    \includegraphics[width=\linewidth]{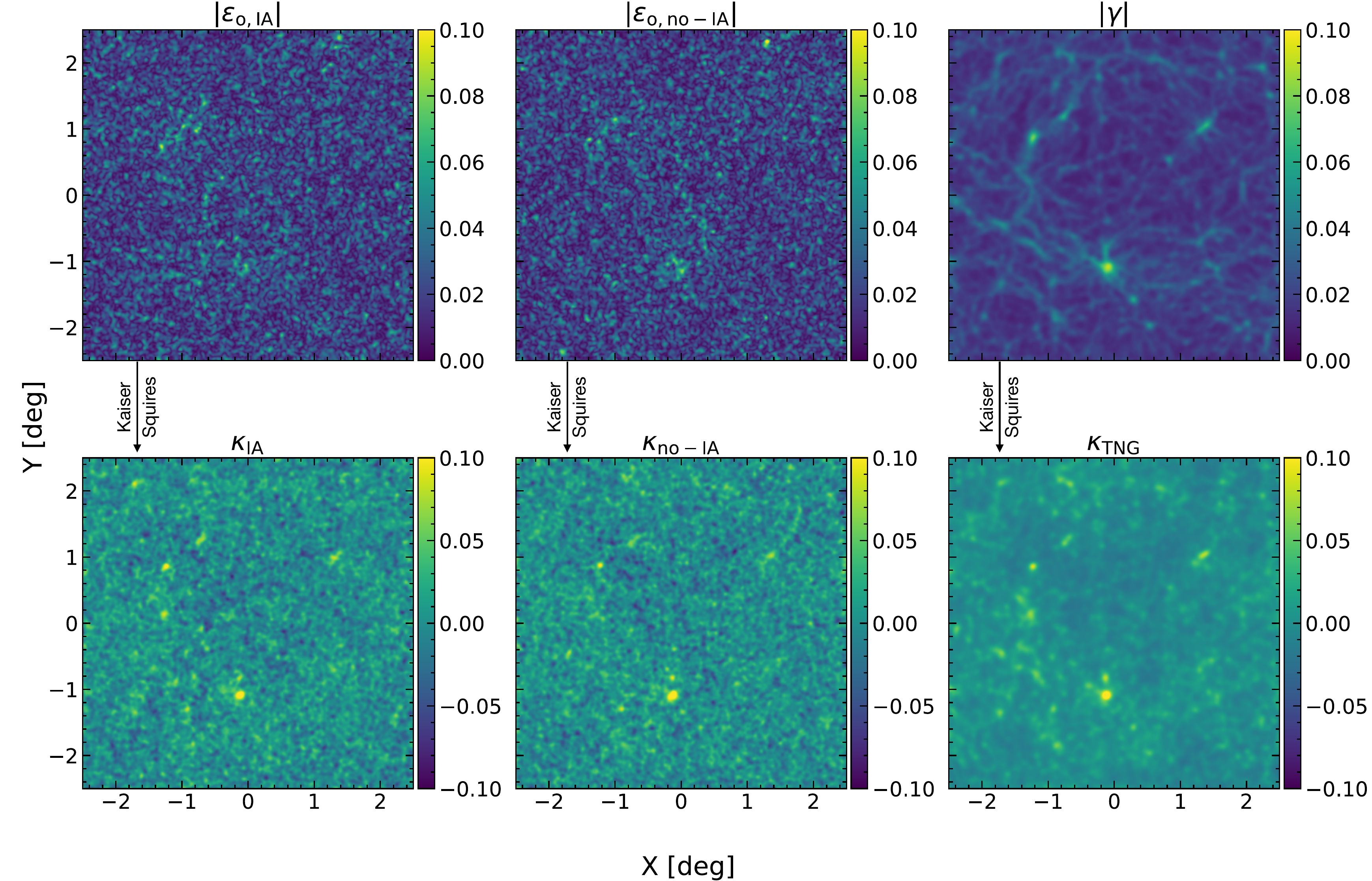}
    \caption{We show examples of typical ellipticity maps and their resulting convergence maps following the procedure of Sec.~\ref{sec:pipeline}. The leftmost column shows the magnitude of observed ellipticity when including the intrinsic shapes of the galaxies, while the center shows the same but when all of the galaxies have been randomly rotated in 3D. The right most pannel shows the shear maps from $\bm{\kappa}$TNG. The bottom pannels show the result of Kaiser-Squires inversion when applied to the observed ellipticity maps and is the resulting convergence maps.}
    \label{fig:ellipt_to_kappa}
\end{figure*}

The angular positions of galaxies inside the light-cone must be aligned with the convergence and shear values generated in the ray-traced $\bm{\kappa}$ and $\bm{\gamma}$ maps. 

To make the galaxy light-cone, we follow the same procedure as generating $\bm{\kappa}$, $\bm{\gamma}$, and $\bm{\beta}$ maps. For each snapshot, we rotate, translate, and project the 3D galaxy positions and luminosity inertia tensors by the same amount and in the same directions as in the lens-plane generations. We are left with the 2D positions and luminosity inertia tensor. We then perform the same random rotations and translations as in the ray tracing of the lens-planes on 2D positions and inertia tensor. Finally, we compute the 2D intrinsic ellipticity after all rotations and translations have been accounted for. 

The final nuance in lining up the galaxies in the TNG300 light-cone with the correct $\bm{\kappa}$ and $\bm{\gamma}$ map coordinates is shifting the galaxies from the true positions to the observed positions, as discussed in the previous section. We compute the total deflection angle between the observer and the $\bm{\kappa}$ map redshifts for each pixel. We then interpolate the positions of the galaxies to find the total deflection for each galaxy. We apply this final shift to the galaxy positions, moving them to the light source's twice-rotated, translated, and deflected angular position. 

Finally, we linearly interpolate the values of $\bm{\kappa}$ and $\bm{\gamma}$ to the observed positions of the galaxies to obtain catalogs of galaxy position, intrinsic ellipticity, convergence, and shear for each galaxy in a given light-cone. 

\begin{figure}
    \centering
    \includegraphics[width=\linewidth]{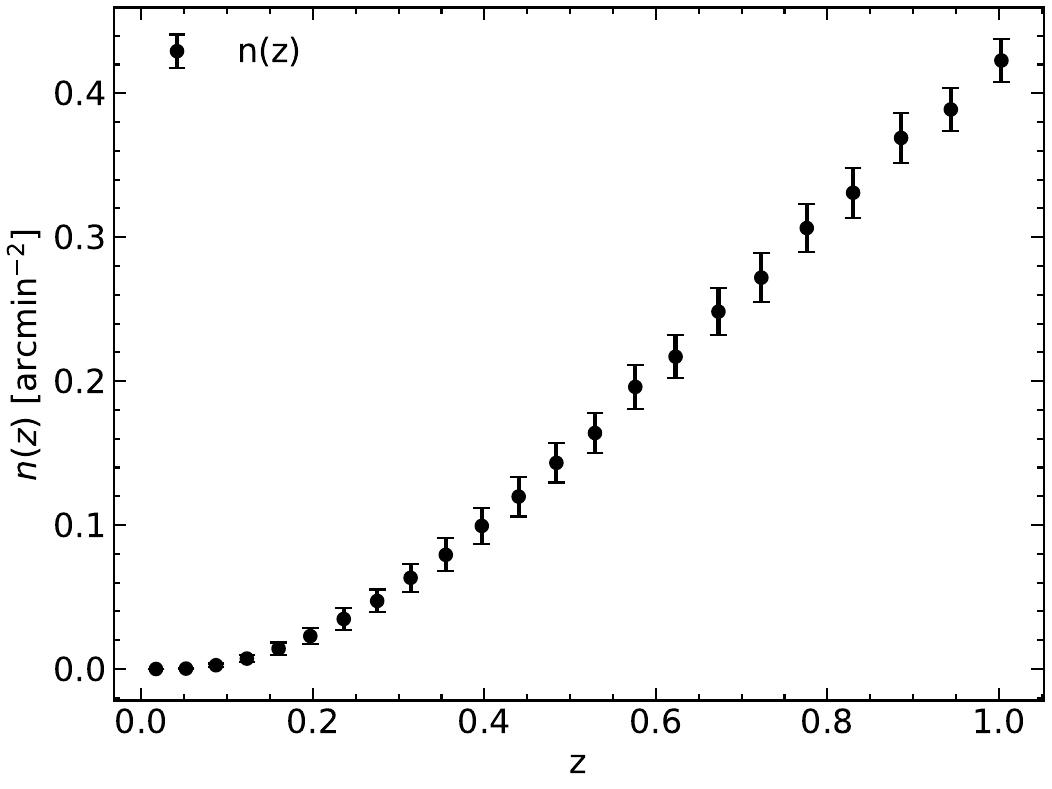}
    \caption{We show the mean and standard deviation of the galaxy redshift distribution over all light-cones used in this work.}
    \label{fig:z_dist}
\end{figure}
We end up with $2,000$ pseudo-independant $25\,\deg^2$ light-cones. In Fig.~\ref{fig:z_dist}, we show the mean galaxy across all light-cones which grows with respect to the redshift. In total, light-cones up to redshift $z=1$ have $\sim 5\,[\rm gal/arcmin^{-2}]$. Again, note that as in the multi-lensplane algorithm, the lens-plane is constructed by stacking snapshots and assuming some redshift evolution, though the galaxy distribution does not change within a given simulation snapshot. We find that this approximation is sufficient to $z=1$, though at higher redshifts, where the number density of galaxies can change drastically within a snapshot one would want to use more and finer sliced snapshots.

\subsection{Creating mock catalogs and maps}
We generate two samples for each of the 2,000 light-cones to compare in this work. The \textbf{IA} samples are generated using the intrinsic shapes of galaxies following Eq.~\ref{eq:ellipticity} exactly. In the \textbf{no-IA case}, random Euler angles are used to apply a 3D rotation to the luminosity inertia tensors before calculating their shapes. This guarantees a random distribution of intrinsic shapes, which, when used with Eq.~\ref{eq:ellipticity}, represents a scenario without intrinsic alignments between the galaxies. 

The result of both the IA and no-IA samples is a catalog of 2,000 pseudo-independent light-cones with each galaxy's position and observed ellipticity. We convert the catalogs of ellipticities to a grid with $1024^2$ pixels where, for each pixel, we compute the weighted ellipticity,
\begin{equation}
\begin{split}
    \langle \bm{\epsilon} \rangle = \dfrac{\sum_j^N W(\theta_j - \theta)\epsilon_j}{\sum_j^NW(\theta_j - \theta)}.
\end{split}
\end{equation}
We use a Gaussian weighting function with a smoothing of $\theta_G = 2.0'$,
\begin{equation}\label{eq:gauss}
    W(\theta) = \dfrac{1}{\pi \theta_G^2}\exp\left[-\dfrac{(\theta - \theta_G)^2}{\theta_G^2} \right].
\end{equation}
\indent The grid of ellipticity is converted to a shear following Eq.~\ref{eq:ellipt_to_gamma} and finally into a convergence map following the Kaiser-Squires inversion, Eq.~\ref{eq:KS}. In Fig.~\ref{fig:ellipt_to_kappa}, we show convergence maps generated following the full process described in this section. The top row shows the magnitude of a single observed ellipticity map at redshift $z_{\rm max}=1$, where the left-most panel uses the true intrinsic shapes of the galaxies in TNG300 (the IA sample), the middle panel uses the randomly rotated galaxy shapes (the no-IA sample), and the right most panel shows the shear with no galaxy shapes included. The bottom panels show the resulting convergence maps after performing Kaiser-Squires inversion. While the convergence level is roughly the same for all three bottom panels, adding the shapes of galaxies introduces a noise component. Interestingly, this causes hotspots in the IA maps, where there are brighter kappa values than in the no-IA case. Similarly, the no-IA case appears to erase some true peaks that would appear in the absence of galaxy shape noise.

\section{Weak Lensing statistics}\label{sec:stats}
We compute an exhaustive list of two-point and non-Gaussian statistics in the following sections. We discuss the metric used to determine statistical distinguishability between the IA and no-IA catalogs, summarize the computation of each of these statistics, and show the results comparing the two catalogs.

\subsection{Statistical Distinguishability}\label{sec:stats1}
We consider the statistical distinguishability of the IA vs. no-IA catalogs following the procedure in \cite{Petri-2016} and \cite{Lee-2023}. This comparison depends on the difference in the expectation value over all realizations of some statistic between the two catalogs $\Delta\langle\mathbf{n}\rangle$, and the covariance matrix of the statistic for one catalog, $\mathbf{C}$. The latter can be defined in either group of maps (though we check that the results are consistent between the two and find that they are), 
\begin{equation}\label{eq:cov}
    C_{ij} = \frac{1}{N-1} \sum_{k=1}^{N}
    \left(n_{i}^{(k)} - \langle n_i\rangle\right)\left(n_{j}^{(k)} - \langle n_j\rangle\right),
\end{equation}
where the sum is over the total number of pseudo-independent realizations (N=2,000), and the subscripts refer to bin numbers in the histogram. To test for statistical distinguishability, we consider the probability with which we can reject the null hypothesis that the IA statistics were drawn from the same distribution as the no-IA (or vice-versa) by computing the $\chi^2$ values of the difference between extracted statistics defined as,
\begin{equation}\label{eq:chi2}
    \chi^2 \equiv {\left(\Delta\langle\mathbf{n}\rangle\right)}^T\, \hat{\mathbf{C}}^{-1}\,\Delta\langle\mathbf{n}\rangle.
\end{equation}
Here $\Delta\langle\mathbf{n}\rangle$ is the mean difference between the IA and no-IA models, and the inverse covariance matrix is given by,
\begin{equation}\label{eq:inv_cov}
    \hat{\mathbf{C}}^{-1} = \frac{N-d-2}{N-1}\mathbf{C}^{-1},
\end{equation}
where the prefactor $(N-d-2)/(N-1)$ is used to debias the precision matrix estimation \citep{hartlap07} and d is the degrees of freedom. For a $\chi^2$ distribution with $20$ degrees of freedom, as is often used in this work, the $1, 2, $ and $3 \sigma$ bounds (corresponding to $\sim 68\%, 95\%$ and $99\%$) are at a $\chi^2=26.22, 34.53, 44.35$ respectively. We consider statistics yielding a $\chi^2$ value above the $3\sigma$ level statistically distinguishable.

In our experiment, we are limited to light-cones of $25\deg^2$, but we can test statistical distinguishability for experiments of larger area, $A$, by scaling the covariance matrix as,
\begin{equation}
    \bm{C}(A) = \left(\dfrac{A}{25\deg^2}\right)^{-1}\bm{C}(25\deg^2).
\end{equation}
We note that we are limited to the galaxy number density provided by our light-cones. Typically, noise is added to simulated convergence maps to account for the intrinsic shapes where this value is inversely proportional to the galaxy number density, but as we are using the distribution of galaxies and their shapes directly from simulations, we are limited to the number of galaxies that exist in our light-cones. We focus on convergence maps with galaxies up to $z_{\rm max}=1$, with a total of $\sim5\,{\rm galaxies}/{\rm arcmin}^2$ (see also Fig.~\ref{fig:z_dist}). Simulation boxes with a higher number density of resolved galaxies should be used in the future for a more detailed comparison, but for this study, we will investigate the effects of survey area scaling at a fixed galaxy number density. 

To present the level of statistical distinguishability between IA and no-IA catalogs for various survey areas, we will show plots with the relative difference between mean IA and no-IA statistics, 
\begin{equation}
\bm{R} = \dfrac{\Delta \langle\bm{n}\rangle}{\bm{n}_{\rm no-IA}}
\end{equation}
and contours of the ratio between the diagonal of the no-IA covariance matrix to the mean of the no-IA catalog as contours,
\begin{equation}\label{eq:contour}
    {\rm Contour} = \sqrt{\dfrac{{\rm diag}(\bm{C}(A))}{ \langle \bm{n}_{\rm no-IA}\rangle^2}}.
\end{equation}
While this excludes the off-diagonal elements of the covariance matrix, it is useful for a quick ``$\chi$-by-eye" analysis of the plots. The true values of $\chi^2$ using the full covariance matrix for each statistic and area can be found in Table~\ref{table:chi_squared}. We use survey areas of the Roman space telescope ($2200\deg^2$)\citep{Roman15} and LSST ($18000\deg^2$)\citep{LSST} for all preceding comparisons. 

We again note here that the comparison to surveys is only in the sense of survey area as we are unable to modify the existing number density. It is further worth noting that we neglect contributions from modes larger than $300\,\rm arcmin$ due to the size of the lightcone, and while all light-cones are pseudo-independant, we are limited in cosmic variance such that we only have a single realization of initial conditions which would be important on large scales. Finally, our covariance matrices are computed following Eq.~\ref{eq:cov}, but we neglect other contributions to the full covariance matrix such as non-Gaussian \citep{takada-2009} and super-sample \citep{Barreira-2018} covariances. Incorpirating these effects would require much larger volume simulations with various initial conditions unavailable to us at this time. We defer the analysis to probe their contributions to a further study.  
\begin{figure*}
    \centering
    \includegraphics[width=\linewidth]{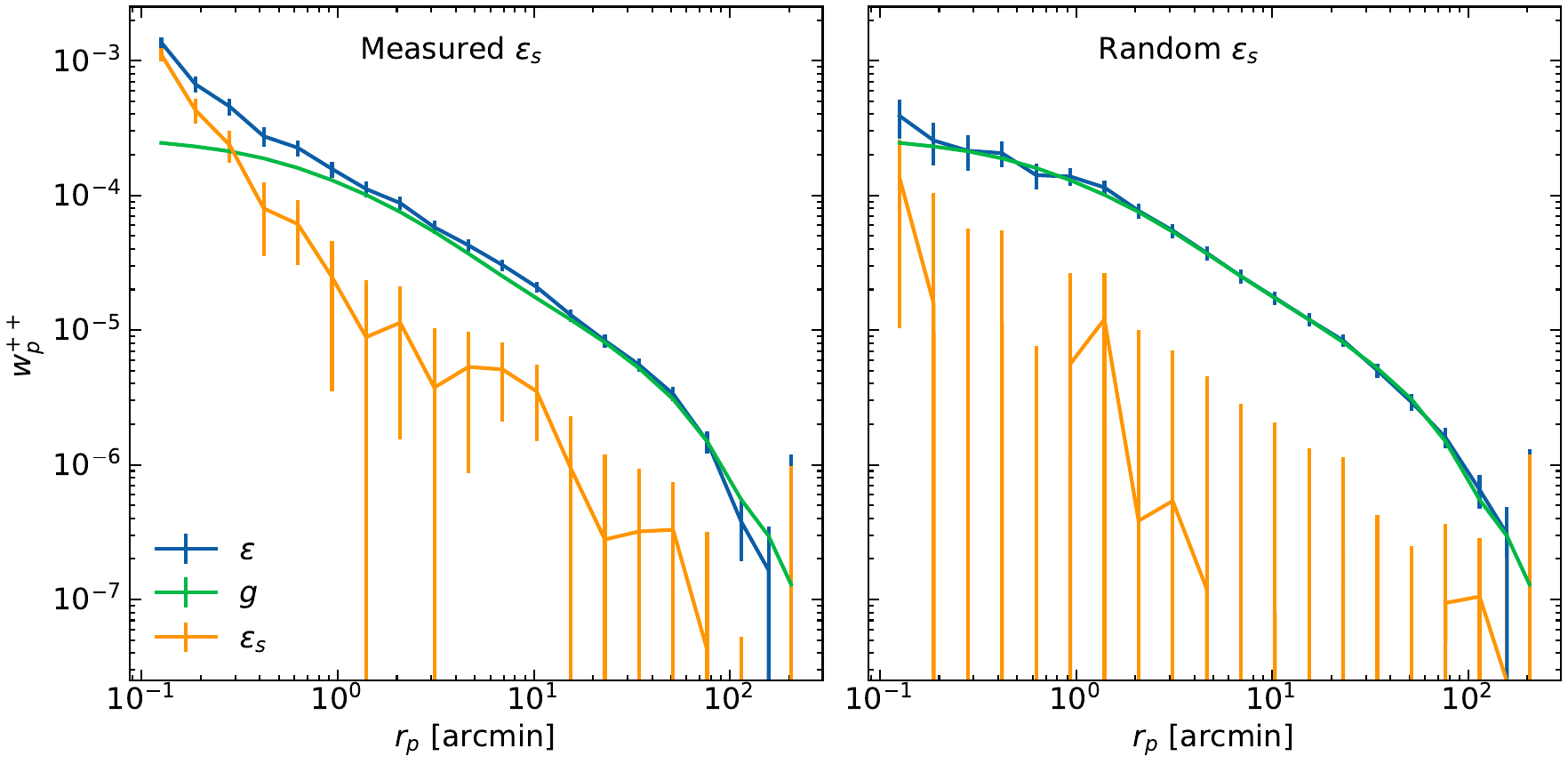}
    \caption{We show the averaged $++$ shape-shape correlation function over all 2,000 light-cones for both the IA catalog (\emph{top panel}) and no-IA (\emph{bottom panel}). In blue is the correlation for the observed ellipticity (Eq.~\ref{eq:ellipticity}), while the intrinsic ($\bm{\epsilon_s}$) and cosmic shear ($\bm{g}$) correlation functions are in orange and green, respectively. As expected, IA enhances the correlation in observed ellipticities, while in the absence of intrinsic alignments, the observed ellipticity traces the large-scale structure shear very closely.}
    \label{fig:corrfunc}
\end{figure*}
\subsection{Correlation function and power spectrum}
Two-point correlations of galaxy shapes are a frequently used statistic in weak lensing analysis. However, intrinsic correlations in the galaxy shapes can bias resulting cosmological parameter estimates \citep{kirk_galaxy_2015, Chisari-2015}. This can be made clear from the definition of the shape-shape correlation function \citep{Lamman-2023}, 
\begin{equation}
    \langle \epsilon^k\epsilon^l\rangle = \langle g^kg^l\rangle + \langle g^k \epsilon_{i}^l \rangle + \langle \epsilon_{i}^k g^l \rangle + \langle \epsilon_{i}^k \epsilon_{i}^l \rangle.
\end{equation}
Here, we have followed the convention of Eq.~\ref{eq:ellipticity} with $\epsilon$ as the observed ellipticity, $\epsilon_s$ the intrinsic ellipticity, and $g$ the reduced cosmic shear. In the absence of intrinsic alignments, the shape-shape correlation function is an unbiased estimator of the shear correlation, but with the presence of intrinsic alignments, the additional terms must be considered. The shape-shape correlation functions for Illustris \citep{Vogelsberger-2014} and MilleniumTNG \citep{delgadoMillenniumTNGProjectIntrinsic2023, Ferlito-2024} galaxies were computed in 
\cite{hilbert_intrinsic_2017} and \cite{delgadoMillenniumTNGProjectIntrinsic2023} respectively, and here we follow a similar analysis as a test of our pipeline and the resulting observed ellipticities. 

We compute the projected $++$ shape-shape correlation functions for both IA and no-IA light-cones, representing correlations both parallel and perpendicular to the vector connecting galaxies. We will call these correlations radial and tangential respectively. We expect tangential alignments to boost positive convergence values to our convergence maps while radial alignments are expected to suppress the convergence. We ignore $\times\times$ correlations which are correlations at a $45\deg$ angle with respect to the vector connecting galaxies. Following \citet{delgadoMillenniumTNGProjectIntrinsic2023}, the 3D correlation function is
\begin{equation}
    \xi^{++}(\bm{r_p}, \bm{\Pi)} = \langle \gamma(\bm{r'}, \hat{\bm{d}})\cdot \gamma(\bm{r'} + \bm{r_p}\hat{\bm{d}} + \bm{\Pi}\hat{\bm{z}}, \hat{\bm{d}}).
\end{equation}
Here, $\bm{\gamma}$ is the real component of the ellipticity of a galaxy in the direction $\hat{\bm{d}}$ pointing to another galaxy, $\hat{\bm{z}}$ is the direction of the line of sight, $\Pi$ is the line of sight distance such that the 3D separation between galaxies is $\bm{r} = (\bm{r_p}^2 + \Pi^2)^{1/2}$, and the expectation value is taken over all $\bm{\hat{r'}}$. The projection is performed by integrating along the line of sight, 
\begin{equation}
    w_p^{++}(\bm{r_p}) = \int_{-\Pi}^\Pi \xi^{++}(\bm{r_p}, \bm{\Pi)}  d\Pi.
\end{equation}
We use the package \textit{Treecorr}\footnote{\href{https://github.com/rmjarvis/TreeCorr}{github.com/rmjarvis/TreeCorr}} \citep{Jarvis-2004} to compute our projected 2D correlation functions with 20 logarithmic radial bins between angles of $0.1$ and $300$ arcmin for all 2,000 light-cones to redshift of $z=1$.

We show the mean correlation function for the observed ellipticity, intrinsic ellipticity, and ellipticity induced by cosmic shearing for both IA (measured $\epsilon_s$) and no-IA (random $\epsilon_s$) scenarios over all light-cones in Fig.~\ref{fig:corrfunc}. The top panel shows that when the true shapes of galaxies are included, the correlation function is boosted by up to nearly an order of magnitude on small scales below a few arcmin. The correlation falls off with respect to distance, becoming consistent with 0 at $\sim 10$ arcmin. This matches expectations from Illustris galaxies in \cite{hilbert_intrinsic_2017}, MilleniumTNG galaxies in \cite{delgadoMillenniumTNGProjectIntrinsic2023}, and has been observed in other simulation suites such as the Horizon-AGN simulations \citep{Chisari-2015}. As the correlation from the intrinsic shapes of galaxies falls off, the observed ellipticity correlation begins to match well the $g$ ellipticity correlation. 

In the case of random ellipticities, we find the expected agreement between the cosmic shear ellipticity and observed ellipticity. This is simply due to the correlations between intrinsic alignments being consistent with zero after random rotations. The enhancement of the correlation at small scales in the IA catalog indicates that either the galaxies tend to align radially or tangentially which can either enhance or suppress the convergence of individual pixels.

We can test this by comparing the angular power spectrum of the convergence map between the IA and no-IA catalogs,
\begin{equation}
    P_l^{kk}(l_i) = \dfrac{1}{N_i}\sum_{|\bm{l}|\in [l_i^{\rm min}, l_i^{\rm max}]} |\bm{\tilde{\kappa}}|^2.
\end{equation}
$N_i$ is the total number of modes, $l_i$ the mean mode in the $i^{th}$ bin, $\tilde{\kappa}$ the Fourier transform of the convergence map, and the sum is over bins in $l$. For each $\bm{\kappa}$ map in the IA and no-IA catalogs, we compute the power spectrum in 20 logarithmically spaced bins between $10\leq l\leq 10000$. As the binning of observed ellipticity includes an effective smoothing of the field with $\theta_G=2\, \rm arcmin$ following Eq.~\ref{eq:gauss}, we perform no additional smoothing, though we exclude pixels within $2\theta_G$ from the edges of each map to remove pixels with incomplete smoothing during pixelization. 
\begin{figure}
    \centering
    \includegraphics[width=\linewidth]{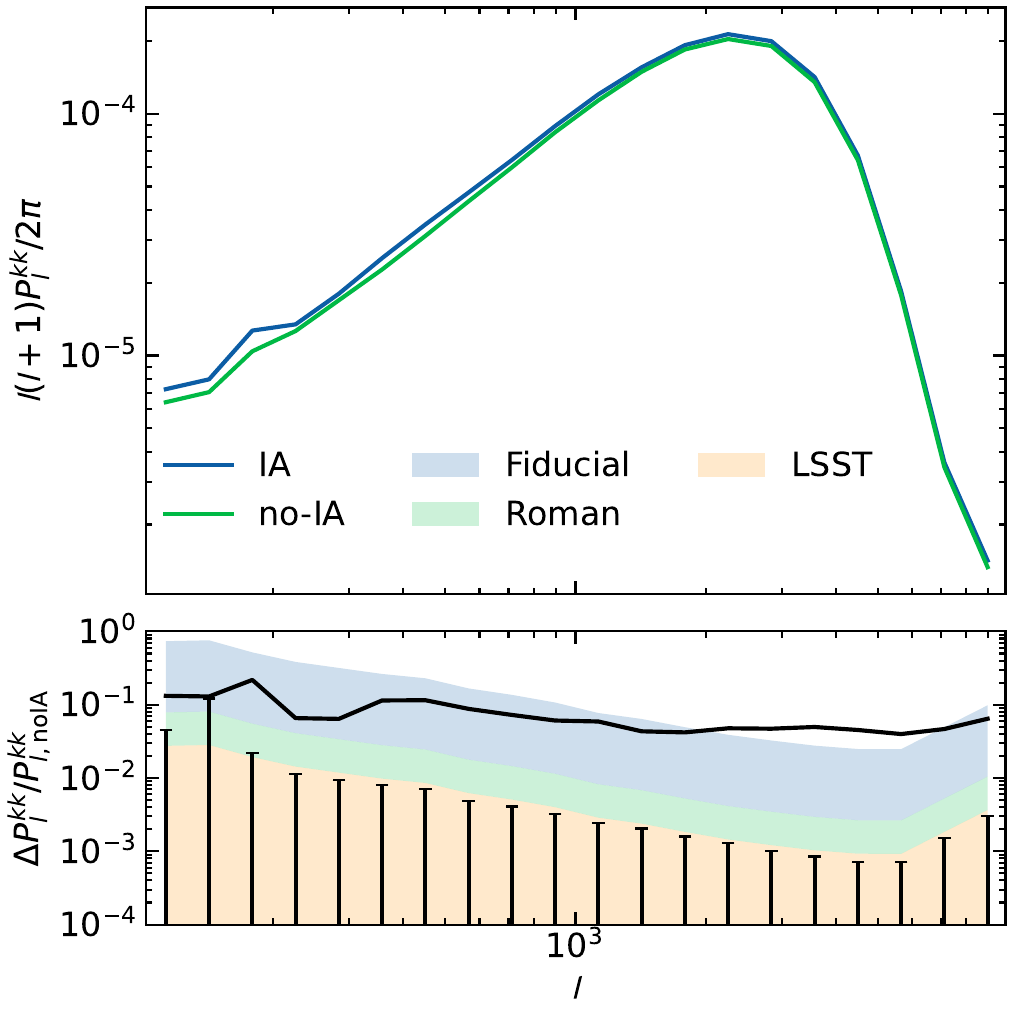}
    \caption{The angular power spectrum averaged over 2,000 $\bm{\kappa}$ map realizations for both the IA and no-IA catalogs. The bottom plot shows the mean relative difference and standard deviation between IA and no-IA power spectra in solid black lines and error bars, respectively. The contours reflect the diagonal of the covariance matrix scaled by the mean power spectrum for surveys of different sizes (see Eq.~\ref{eq:contour}).}
    \label{fig:powerspec}
\end{figure}
\begin{figure*}
    \centering
    \includegraphics[width=\linewidth]{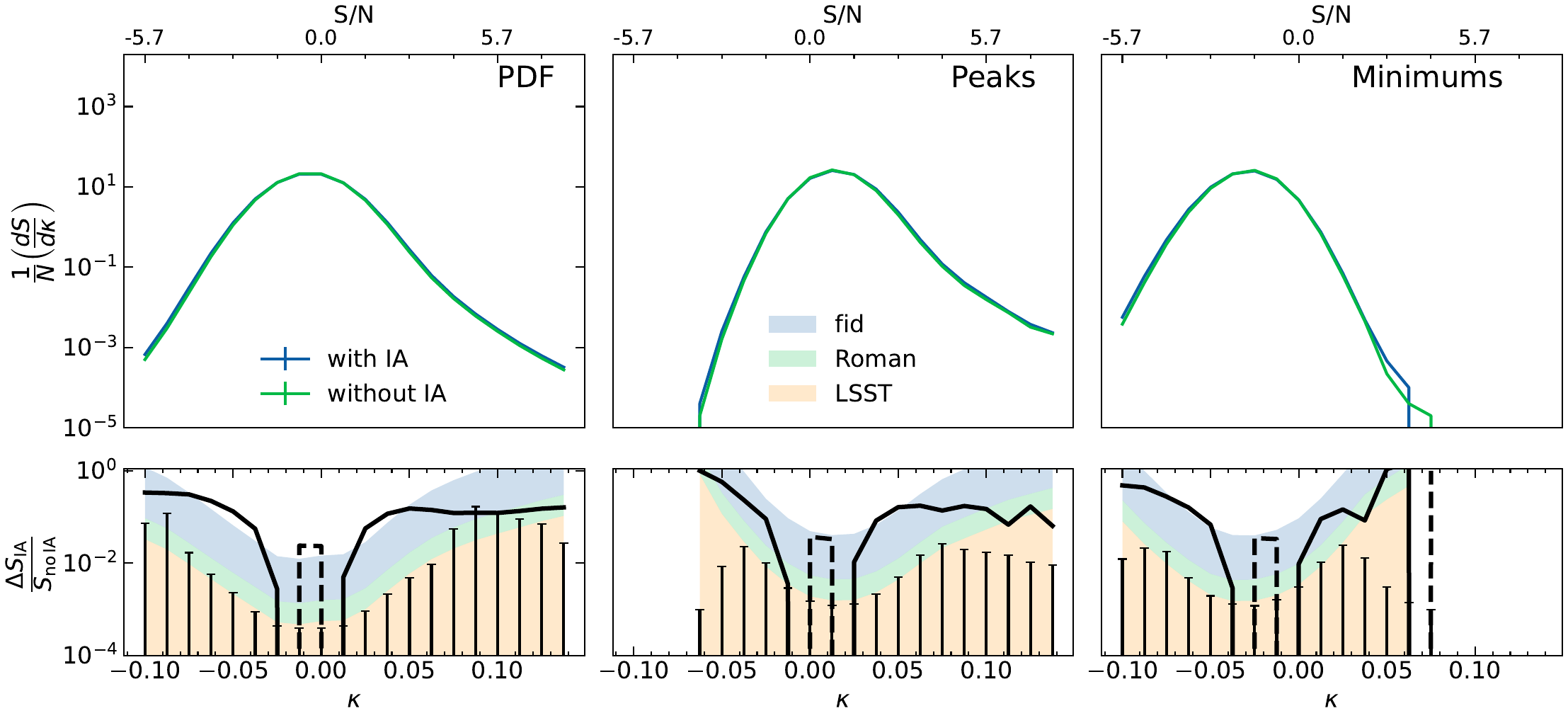}
    \caption{The averaged one-point probability, peak, and minimum distribution functions for the IA and no-IA convergence map catalogs are shown with 20 bins between $-0.1\leq\kappa\leq0.15$. The lower panels show the mean and standard deviation of the relative difference for each of the statistics along with contours similar to those in Fig.~\ref{fig:powerspec}. The dashed line represents negative values of the relative difference, which are initially hidden due to the logarithmic axis. We see that IA provides a noticeable broadening to the $\bm{\kappa}$ distribution enhancing both intermediate significance peaks and minima, and that this broadening is statistically distinguishable for surveys larger than the fiducial $25\deg^2$ survey.}
    \label{fig:stats}
\end{figure*}

In Fig.~\ref{fig:powerspec}, we show the angular power spectrum of the convergence maps from the IA and no-IA catalogs. The top panel shows both catalogs' mean scaled power spectrum, while the bottom panel presents the relative difference. The error bars at the bottom are the standard deviation of the relative difference over the $2,000$ maps. The fact that the solid line does not cross through any of the error bars implies that the relative difference is inconsistent with $0$. There is a clear boost in power due to IA, as predicted from Fig.~\ref{fig:corrfunc}. Further, we see that the boost in convergence is across all scales, with a maximum of $\sim 10\%$ at the largest scales, going down to the percent level at smaller scales of $l>1000$. Following Eq.~\ref{eq:contour}, we show the level of statistical distinguishability for the fiducial $25\deg^2$ boxes, as well as for a \textit{Roman}, and LSST-like survey. For the fiducial $25\deg^2$ survey, the two catalogs are virtually indistinguishable with a $\chi^2=5.2$, but for larger surveys, the IA vs no-IA scenarios are far beyond the $3\sigma$ level of distinguishability. We list the values in Table~\ref{table:chi_squared}.

The angular power spectrum and correlation function results are consistent in that the enhanced correlations due to IA lead to a boost in power at the convergence map level implying that there an overall tangential correlation due to IA. The results presented so far do not indicate which environments are affected; rather, they indicate only that the convergence is boosted on all scales. In the following sections, we explore enhanced kappa values due to IA at the pixel level, which offers insight into how we expect specific environments to be affected by IA.  
\subsection{Pixel level PDF, peak and minimum counts}
We compute each convergence map's one-point probability (PDF), peak count, and minimum distribution functions. Each statistic has been shown to significantly tighten the errors on cosmological constraints compared to two-point functions alone \citep{Coulton-2020, thiele_accurate_2020, Osato-2021, Lu-2023}.

The PDF is computed using all $1024^2$ pixels in a given map, while for peaks (minima), we find the values of pixels with higher (lower) $\bm{\kappa}$ values compared to their eight nearest neighbors. For each distribution function, we use 20 linearly spaced $\bm{\kappa}$ bins between $-0.1\leq\kappa\leq 0.15$ for consistency, though, for peaks and minima, there are empty bins due to the lack of low $\bm{\kappa}$ peaks and high $\bm{\kappa}$ minima. For peaks, we then have 17 bins between $-0.05\leq \bm{\kappa}\leq0.15$, and for minima, there are 12 bins between $-0.1\leq\kappa\leq0.17$. 

In Fig~\ref{fig:stats}, we present the distribution functions for the PDF, peaks, and minima in both the IA and no-IA cases, respectively. The bottom panels are shown similarly to Fig.~\ref{fig:powerspec}, where the line represents the relative difference between the histograms, the errorbars show the standard deviation of the relative difference over the 2,000 maps, and the contours follow Eq.~\ref{eq:contour} for the fiducial $25\deg^2$ box and the areas corresponding to the three other surveys of interest. 

We also show the commonly used signal-to-noise ratio, $S/N$, as a second x-axis, where $N=\sqrt{\langle\kappa^2\rangle}$, or the root mean square (RMS) of $\bm{\kappa}$. For consistency, we compute the RMS of all $\bm{\kappa}$ in the no-IA catalog and use the mean for all computations. This quantity represents the significance of a peak or minimum and is a proxy for the effective mass of the lens that created it. The largest $S/N$ peaks are thought to be generated by a single massive lens, lower $S/N$ peaks are thought to arise from multiple distortions of lower-mass lenses, and peaks with negative $S/N$ are thought to arise around void regions \citep{Yang-2011, liu_origin_2016}.

For each statistic, the comparisons between IA and no-IA are relatively indistinguishable for the fiducial survey area, similar to the power spectrum in Fig.~\ref{fig:powerspec}. All of the statistics become distinguishable as the survey increases in area. We present the values of $\chi^2$ from Eq.~\ref{eq:chi2} in Table.~\ref{table:chi_squared}.

There is a clear difference between the IA and no-IA catalogs at the PDF level, which is reflected in the peak and minimum count histograms. In the presence of IA, kappa distributions are broadened to enhance the high and low-value pixels, while the intermediate values around $\kappa\sim0$ are suppressed. The results in Fig.~\ref{fig:stats} reflect the angular power spectrum and correlation function results where we find an overall enhancement to the convergence values, but this effect appears to be in both regions of particularly large positive and of particularly large negative convergence, such as around galaxy groups and clusters and around void regions, respectively. For the galaxy group to cluster lenses to experience a boost, the alignments must be primarily tangential, while for void regions to become more negative, the regions must have primarily radial alignments. A more in-depth study of the connection between the origin of weak lensing peaks and minima and their connection to the intrinsic alignments of galaxies is required to disentangle this result fully, which is beyond the scope of this paper but a plan for future work. 

\subsection{Minkowski Functionals}
\begin{figure*}
    \centering
    \includegraphics[width=\linewidth]{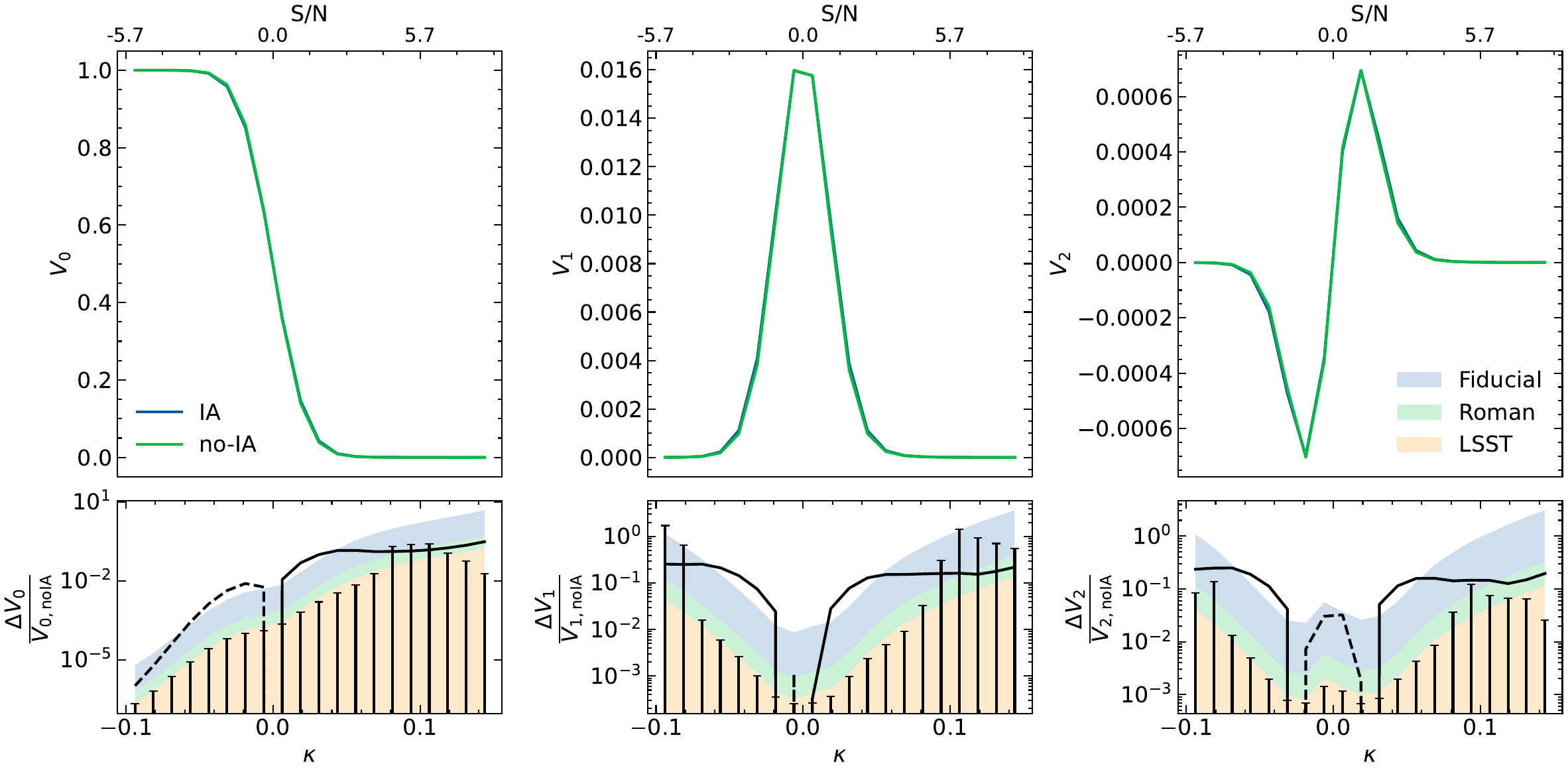}
    \caption{The same as Fig.~\ref{fig:powerspec} and Fig.~\ref{fig:stats}, but for each of the Minkowski functionals described in Eq.~\ref{eq:minkowski}.}
    \label{fig:minkowski}
\end{figure*}
Minkowski functionals describe the geometric properties of some mathematical space \citep{petri_cosmology_2013}. In weak lensing, the geometrical space is described by pixels in some thresholded region given by the excursion sets $\Sigma(\nu) = \{\bm{\kappa}> \nu\}$, where $\nu$ represents some threshold values of the field. Minkowski functionals are a powerful descriptor of the weak lensing field as they encode arbitrarily high-order correlation functions, making them a powerful tool to probe non-Gaussianities, particularly at small scales \citep{kratochvil_probing_2012}. For a 2D field, there are three Minkowski functionals available, $V_0$, $V_1$, and $V_2$,
\begin{equation}\label{eq:minkowski}
\begin{split}
    V_0(\nu) &= \dfrac{1}{A}\int_{\Sigma(\nu)}da\\
    V_1(\nu) &= \dfrac{1}{4A}\int_{\partial\Sigma(\nu)}dl\\
    V_2(\nu) &= \dfrac{1}{2\pi A}\int_{\Sigma(\nu)}\mathcal{K}dl.\\
\end{split}
\end{equation}
$V_0$ can be considered the fractional area of pixels above a given threshold $\nu$ with respect to the total area $A$. $V_1$ is the size of the boundaries for each threshold, and $V_2$ is the characteristic shape of the regions above the threshold with $\mathcal{K}$ as the curvature. We refer the reader to \citet{kratochvil_probing_2012} and \citet{petri_cosmology_2013} for an in-depth discussion of cosmological applications and derivations of Minkowski functionals and to \citet{armijo_cosmological_2025} for recent applications to HyperSuprime Cam weak lensing observations.  

We generate a linearly spaced thresholding array in $\bm{\kappa}$ with $20$ values between $-0.1\leq\kappa\leq0.15$. We use the \textit{Lenstools}\footnote{\href{https://lenstools.readthedocs.io/en/latest/}{lenstools.readthedocs.io}} \citep{Petri-2016_lenstools} package to compute the Minkowski functionals and compare the mean of the IA and no-IA catalogs, similar to the comparisons made in the previous sections. 

We present the comparison in Fig.~\ref{fig:minkowski}, where the top shows the averaged Minkowski functionals over the 2,000 convergence maps, while the bottom shows the relative difference. Again we find that the results are mostly statistically distinguishable for large survey areas, though they are noisier than the previous statistics studied. For the edges in $\bm{\kappa}$, we see that the statistical fluctuations in the relative difference are large, meaning they are consistent. However, the trends are consistent with the previous statistics studied. The IA enhances kappa particularly in high and low $\bm{\kappa}$ regions, leading to a boost in $V_0$ at high kappa and suppression at low kappa (as $V_0$ is essentially the cumulative distribution function). Similarly, the length of the excursion sets in these regions are boosted (given by $V_1$) and the shapes become more complex (given by $V_2$). We again list the $\chi^2$ for each statistic in Table.~\ref{table:chi_squared}.

\section{Discussion \& Conclusion}\label{sec:conclusion}
\begin{table}
\centering
\begin{tabular}{lcccc}
\hline
\textbf{Statistic} & $\bm{\chi^2}_{\rm D.o.F.}({\rm Fid})$ & $\bm{\chi^2}_{\rm D.o.F.}({\rm Roman})$ & $\bm{\chi^2}_{\rm D.o.F.}({\rm LSST})$ \\
\hline
\(P_l^{kk}\) & \(0.26\) & \(22.88\)  & \(187.22\) \\
PDF     & \(0.18\)      & \(16.22\)   & \(132.74\) \\
Peaks   & \(0.20\)      & \(17.61\)   & \(144.07\) \\
minima  & \(0.20\)      & \(18.28\)   & \(149.62\) \\
\(V_0\) & \(0.18\)      & \(16.12\)  & \(131.90\) \\
\(V_1\) & \(0.27\)      & \(23.99\)   & \(196.32\) \\
\(V_2\) & \(0.22\)      & \(19.54\)  & \(159.95\) \\
\hline
\end{tabular}
\caption{\(\chi^2_{\rm D.o.F.}\) values for the various weak lensing statistics and survey areas studied in this work. We compute the statistics for various survey areas corresponding to existing and upcoming weak lensing surveys. We find that IA and no-IA are statistically indistinguishable for the fiducial area of this work, but for larger area surveys, the statistics become distinguishable above the $3\sigma$ level.}
\label{table:chi_squared}
\end{table}

In this work, we explore how the intrinsic alignments of galaxies drawn directly from hydrodynamical simulations impact various statistics of the weak lensing field. We extract galaxy shapes from the TNG300 simulations and include them in the ray-tracing pipeline to generate $2,000$ pseudo-independent weak lensing convergence maps. This set of maps is compared to convergence maps generated with randomly rotated galaxy shapes corresponding to no intrinsic alignments. We then compare the IA and no-IA catalogs using two-point statistics, including the shear-shear correlation function and angular power spectrum, as well as non-Gaussian statistics, such as the one-point, peak, minimum distribution functions, and three Minkowski functionals.

For each statistic, we compute the level of statistical distinguishability between the IA and no-IA scenarios using a $\chi^2$ statistic following Eq.~\ref{eq:chi2} to determine at what survey area the effects would be noticeable and potentially lead to a bias in cosmological constraints. We list the values of $\chi^2$ for each test and survey size in Table~\ref{table:chi_squared}. We find that for all statistics considered, surveys with areas above the fiducial $25\deg^2$ show above $3\sigma$ levels of statistical distinguishability between IA and no-IA catalogs. This is strong evidence that IA can induce a bias in cosmological parameter constraints from both two-point and non-Gaussian statistics and should be modeled correctly in analyses for ongoing and future surveys.

Another interesting result from this work is that we find IA to broaden the convergence distribution, implying that IA induces a positive convergence signal in high-density regions and negative convergence in low-density regions. In contrast, at intermediate values of $\bm{\kappa}$, the signal is suppressed relative to random galaxy alignments. This result indicates environmental dependence of IA. For lensing in the most overdense regions, there must be a high level of tangential alignment. In contrast, galaxies are preferentially radially aligned in void-like regions, causing more negative convergence values. It would be valuable to investigate this environmental dependence originating from IA in future work, and models should also account for such effects. In future work, we also plan to compare the effects of the various IA models, such as the linear alignment, non-linear alignment, and tidal torquing models, with IA drawn directly from hydrodynamical simulations. 

We caution that as mentioned in \S\ref{sec:stats1}, this study does not include any survey noise, non-Gaussian or super-sample covariance, so we are not able to conclude that surveys with areas above $\sim 25\deg^2$ will inherently find an IA signal. Further, the light-cones used in this work are limited in galaxy number density compared to upcoming surveys such as LSST and drawn all from the same initial conditions, which could affect the large-scale modes. It would be interesting to see how IA changes in simulations with higher galaxy number densities, larger volumes, and various initial conditions.

\section*{Acknowledgements}

We thank Fulvio Ferlito, Ana Maria Delgado, and Ken Osato for helpful conversations during this work. MEL is supported by NSF grant DGE-2036197. ZH acknowledges financial support from NASA ATP grant 80NSSC24K1093. The Flatiron Institute is supported by the Simons Foundation.

\bibliography{biblio}{}
\bibliographystyle{aasjournal}
\appendix
\section{Rotations in ray-tracing}\label{sec:flip}
The order of operations in the ray-tracing algorithm is important. In some ray-tracing algorithms, such as those used here in \S~\ref{sec:ray-tracing} and used to generate $\bm{\kappa}$TNG, the 3D matter distribution is randomly rotated and projected to a 2D overdensity field. The derivatives of the overdensity field are then computed to generate lens-planes. After saving these lens-planes, pseudo-independent weak lensing maps can be generated with random rotations and translations of the pre-saved lens-planes \citep{Petri-2016, Osato-2021}. 

Following this order of operations can generate problematic shear values when rotations are made by $90\deg$ or $270\deg$. This issue arises because the lens planes are computed using derivatives in an unrotated coordinate system but are later used in a rotated system. Specifically, rotations of $90^\circ$ or $270^\circ$ introduce a multiplicative factor of $-1$ to the shear values. The total shear at a given redshift depends on the shear values at previous redshifts, as seen in Eq.~\ref{eq:beta}. So, the resulting total shear when the lens-planes have all been rotated randomly is not simply inverted by $-1$, but a combination of weighted shears, some of which have been assigned a factor of $-1$ (if rotated by $90\deg$ or $270\deg$) while others have a factor of $+1$  (if rotated by $0\deg$ or $180\deg$). Care must be taken to either rotate only by 0 or 180 degrees, to compute the lens-planes after coordinate system rotations, or to correct the shear maps that have been rotated to problematic angles. 

Further, the direction of ray-tracing matters for shear. For a ray starting at the observer (often called backward ray-tracing), the projected coordinate system is rotated by $180$ degrees about the projection axis compared to a ray originating at the source. This introduces another problematic multiplicative factor of $-1$ into the shear of galaxies. Because of this effect, if backward ray-tracing, the lens-planes must first be rotated by $180^\circ$ about the projection axis to remove the $-1$ introduced from the ray-tracing direction.

In this work, we have taken a conservative approach. We limit the random rotations of our lens-planes to $90\deg$ and $270\deg$, which simultaneously accounts for both projection and coordinate transformation issues. We check that our shears values are sensible and that Kaiser-Squires inversion of our ray-traced $\bm{\gamma}$ maps produce convergence maps that match those generated from ray-tracing.



\end{document}